\documentclass[runningheads]{llncs}
\usepackage[T1]{fontenc}
\usepackage{graphicx}
\usepackage{amsmath,amssymb}
\usepackage{hyperref}
\usepackage{float}
\usepackage{comment}
\usepackage{orcidlink}
\usepackage{bbding}
\usepackage{subcaption}
\usepackage{placeins}

\title{Validating Multi-Model AI-Generated Acceptance Criteria in Requirements Engineering}

\titlerunning{Validating AI-Generated Acceptance Criteria}

\author{Mateen Ahmed Abbasi\inst{1} \orcidlink{0000-0001-8988-8816}\,\Envelope \and
Petri Ihantola\inst{1} \orcidlink{0000-0003-1197-7266} \and
Tommi Mikkonen\inst{1} \orcidlink{0000-0002-8540-9918} \and
Niko Mäkitalo\inst{1}\orcidlink{0000-0002-7994-3700} }

\authorrunning{M. A. Abbasi et al.}

\institute{Faculty of Information Technology, University of Jyväskylä, Jyväskylä, Finland\\
\email{\{mateen.a.abbasi, petri.j.ihantola, tommi.j.mikkonen, niko.k.makitalo\}@jyu.fi}}

\begin{document}

\maketitle

\begin{abstract}
Large language models (LLMs) are increasingly used in software engineering to generate artifacts such as requirements, user stories, acceptance criteria, documentation, and code. Their use in requirements engineering raises important questions about correctness, traceability, and hallucination, especially because flawed specifications can affect later design, implementation, and verification activities. While much recent work on AI-assisted software engineering has focused on code generation and code-level validation, the validation of AI-generated specifications has received less attention. This paper addresses this gap by presenting a validation-oriented approach for AI-generated acceptance criteria. The approach combines retrieval-augmented generation (RAG), multi-model generation, and an explicitly structured human-in-the-loop workflow. We describe a working prototype that allows users to prepare or enter user stories, generate acceptance criteria using several LLMs, optionally augment generation with retrieved requirements-related examples, compare outputs side by side, inspect retrieved context, and select, edit, or regenerate candidate criteria before explicitly submitting a reviewer-finalized version. The system also records model outputs, user interventions, retrieval settings, and final criteria as part of a logged decision process. Rather than claiming that generation accuracy is already improved, the contribution of this work is to operationalize validation as an explicit, inspectable, and accountable process for AI-generated requirements artifacts.

\end{abstract}

\section{Introduction}

The use of large language models (LLMs) in the context of software engineering has expanded rapidly, ranging from code generation to analysis and documentation \cite{hou2024large}. The integration of LLMs into RE processes offers a promising roadmap for addressing long-standing challenges in requirements elicitation, analysis, specification, and validation \cite{arora2024advancing}. In requirements engineering, LLMs have been studied as tools to support the creation of requirements, user stories, acceptance criteria, and other early artifacts \cite{cheng2026generative}. However, these capabilities raise questions about the correctness and reliability of AI-generated specifications. 

Previous research has explored task-specific instruction tuning for requirements generation \cite{habib2025reqbrain}. Nevertheless, the ability to produce plausible requirements-related text does not establish its correctness for a particular project. Acceptance criteria describe conditions used to assess whether the expected system behavior satisfies a requirement or user story. Incorrect, incomplete, or unsupported acceptance criteria can therefore affect subsequent implementation and testing activities.

The central challenge is therefore not only whether AI can generate acceptance criteria, but how to validate such criteria in a transparent, accountable, and practical way. In this work, validation is treated as a first-class activity rather than as an informal correction stage after generation. The proposed approach combines optional retrieval augmentation, multi-model comparison, and explicit human ratification to support the inspection and revision of AI-generated acceptance criteria before finalization. By exposing retrieved context and recording human decisions, the workflow makes the generation and validation process more transparent and amenable to subsequent analysis. This framing is particularly relevant in RE, where early specification artifacts often require interpretation, negotiation, and contextual judgment rather than purely formal proof.

This work continues our ongoing research on human-AI collaboration in requirements engineering. In earlier work, we examined how AI is reshaping RE practices and highlighted persistent concerns around ambiguity, traceability, transparency, and trustworthy human-AI collaboration \cite{abbasi2025reconsidering}. Later, based on the analysis of the existing literature, we proposed the Human-AI Requirements Engineering Synergy Model (HARE-SM), which frames AI-supported RE as a collaborative process requiring human oversight, explainability, and feedback loops \cite{abbasi2025towards}. This paper moves from the broader conceptual framing towards a concrete validation-oriented prototype. Specifically, it focuses on AI-generated acceptance criteria and investigates how retrieval grounding, multi-model comparison, and explicit human ratification can support validation before generated criteria become final.

 This paper makes three contributions:
\begin{itemize}
    
    \item It proposes a hybrid workflow for generating and validating acceptance criteria from user stories by combining RAG with explicit human-in-the-loop (HITL) validation.
   
    \item It presents a working multi-model prototype that supports optional retrieval, side-by-side comparison of acceptance criteria produced by different language models, and structured user actions for accepting, editing, or regenerating outputs.

   \item It introduces a logging design that records generated outputs, user interventions, retrieval settings, and reviewer-finalized criteria, thereby turning validation into an analysis-ready process rather than an informal post-hoc correction step.
\end{itemize}
These contributions position the prototype not as a finished accuracy solution, but as an experimental platform for studying the validation of AI-generated requirements artifacts in a transparent and accountable way.

\section{Background and Motivation}

\subsection{AI Support for RE}

Natural language processing (NLP) has been widely used in RE to support tasks such as classification, traceability, and quality analysis \cite{zhao2021natural}. Recent advances in LLMs have sparked AI-assisted requirements generation, and recent reviews document growing interest in LLM-supported RE activities \cite{zadenoori2025large}. At the same time, recent research identifies limitations and risks in LLM-supported requirements engineering, including hallucinated content, limited interpretability, and challenges in evaluating generated artifacts \cite{cheng2026generative}. Unlike traditional NLP-based approaches, which tend to classify or detect features in already existing requirements, generative models produce new textual artifacts that may appear fluent and correct but actually lack support or may even conflict with other aspects of the project. This is particularly important for acceptance criteria. Acceptance criteria are intended to make the conditions for satisfying a requirement or user story more explicit and verifiable. However, generated criteria may still omit important constraints or introduce unsupported assumptions. Human validation is therefore needed to assess their alignment with the intended requirement and relevant project context.

For this reason, the use of LLMs in RE should not be understood only as a generation problem. It is also a validation problem. The generated artifact must be checked for correctness, grounding, traceability, and alignment with stakeholder intent before it can be treated as a reliable requirement artifact.

\subsection{From Generation to Validation}

Recent work on generative AI for RE shows that LLMs are increasingly applied to language-intensive RE tasks, but also highlights concerns related to explainability, reproducibility, controllability, and evaluation~\cite{cheng2026generative}. These concerns are important because requirements artifacts are not only technical descriptions; they also reflect stakeholder needs, domain assumptions, and project decisions. A generated requirement or acceptance criterion may be syntactically well formed while still being wrong from the viewpoint of the project.

Prior work has also investigated the generation of acceptance criteria from user stories. For example, Li et al. proposed SimAC, an approach that uses LLM-based simulated agile collaboration to generate acceptance criteria \cite{li2024simac}. While approaches such as SimAC primarily investigate the generation of acceptance criteria, our work emphasizes the subsequent human validation process. Specifically, we operationalize multi-model comparison, inspection of retrieved context, reviewer interventions, and explicit finalization within a single instrumented workflow. Validation in this context means that candidate criteria should be reviewed against the user story, available project context, and human judgment. This position is consistent with the broader view of RE, where validation often depends on stakeholder assessment, interpretation, and agreement rather than only formal checking~\cite{nuseibeh2000requirements}.

\subsection{Fine-Tuning and Retrieval for Validation}

Fine-tuning can adapt model behavior to a particular task or domain. Previous research has explored task-specific instruction tuning for requirements generation \cite{habib2025reqbrain}. However, fine-tuning does not by itself expose which contextual sources support an individual generated criterion. Retrieval provides a different mechanism by supplying external information as part of the generation context \cite{lewis2020retrieval}. These approaches are not mutually exclusive.

In the current prototype, retrieval was selected to make the contextual information supplied during generation available for human inspection. This design allows reviewers to examine the retrieved material when assessing candidate acceptance criteria. However, the availability of retrieved context does not by itself establish that the generated criteria are correct or supported by the requirements of a particular project.

\subsection{Retrieval-Augmented Generation and Traceability}

Retrieval-Augmented Generation (RAG) offers a practical way to incorporate project-specific information into LLM prompts without changing the model weights~\cite{lewis2020retrieval}. In RE, this is useful because requirements artifacts often need to be interpreted in relation to other documents, such as project descriptions, specifications, stakeholder notes, or previous requirements. By retrieving and exposing relevant context, RAG can support grounding and traceability during validation.

Recent RE-related work has already explored RAG for tasks such as generating test scenarios from natural-language requirements and recovering traceability links~\cite{arora2024generating,hey2025requirements}. These studies show that retrieval-based methods are relevant for connecting generated or analyzed artifacts with supporting textual evidence. The present work follows this direction but focuses specifically on acceptance criteria and on the human validation process around AI-generated criteria.

\subsection{Human-in-the-loop Validation and Explainability}

In RE, human reviewers must remain responsible for approving requirements content. Human-in-the-loop workflows are therefore important because they keep the human reviewer involved in judging the adequacy of generated artifacts. In AI-supported work, explainability, user confirmation, and iterative refinement are important factors for acceptance and responsible use~\cite{theis2023requirements}. General human--AI interaction guidelines also emphasize supporting user correction and appropriate control over AI-assisted activities \cite{amershi2019guidelines}.

In our context, HITL is designed as a central part of the validation workflow rather than as an informal correction stage. Users can compare outputs when multiple models are selected, inspect available retrieval context, and select or revise candidate acceptance criteria. The final result is explicitly submitted by the reviewer, and the corresponding interaction events are recorded. Our earlier HARE-SM framework similarly emphasized that AI-supported RE should combine automation with human oversight, explainability, and iterative feedback~\cite{abbasi2025towards}.

\section{Overview of the Proposed Approach}

The process starts with a user story or an informal requirement that is converted into a user story. The prototype supports three input modes: users may paste an existing user story, construct one through a guided form, or provide a plain-language requirement that the system converts into candidate user stories. This design supports participants with different levels of requirements-engineering expertise while keeping the downstream validation workflow consistent.

Optionally, the user enables retrieval, which allows the system to retrieve relevant material from a collection of requirements-related examples. The retrieved text is represented using TF--IDF and ranked by cosine similarity. The highest-ranked chunks are added to the generation prompt for the selected model when retrieval is enabled, following the general principles of retrieval-augmented generation \cite{lewis2020retrieval}.

The generated acceptance criteria are then displayed to the user. The system does not presume that automatically generated output is correct. Instead, users compare the outputs, inspect retrieved context when available, select or edit relevant criteria, and explicitly decide whether to accept, edit, or regenerate the generated content. All user decisions are explicitly recorded, and none of the artifacts are treated as final until they are ratified by the user.

\begin{figure}[!h]
    \centering
        \includegraphics[
        width=1.0\linewidth,
        height=1.0\textheight,
        keepaspectratio]{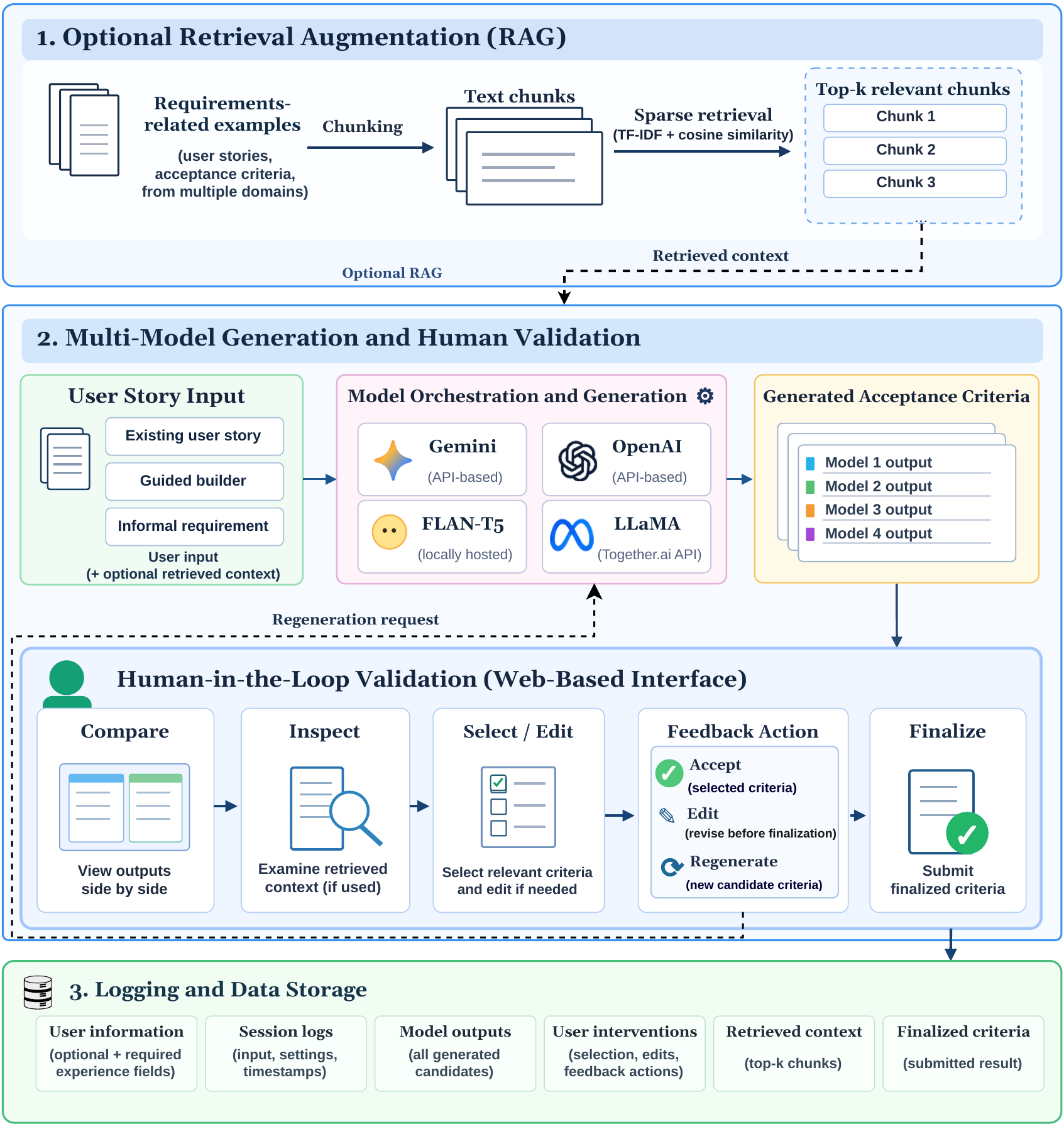}
    \caption{Architecture of the Acceptance Criteria Assistant,
    showing optional retrieval augmentation, multi-model
    generation, human validation, and interaction logging.}
\label{fig:HITLRAG}
\end{figure}

Figure~\ref{fig:HITLRAG} presents the end-to-end architecture, combining two concerns: \textbf{(i) optional retrieval} of requirements-related examples and \textbf{(ii) human-in-the-loop validation through the interface.}

\section{Methodology: Retrieval-Augmented Generation for RE}

\subsection{Corpus Preparation and Chunking}
The retrieval corpus consists of cross-domain example user stories and associated acceptance-criteria material. The text is divided into chunks of approximately 200 FLAN-T5 tokens. The chunks are indexed for sparse retrieval and used as additional context when retrieval is enabled. The corpus provides requirements-related examples rather than authoritative documentation for each participant's software project.

\subsection{Sparse Retrieval with TF-IDF and Cosine Similarity}
For retrieval, we use a lightweight sparse TF--IDF vectorization approach and rank candidate chunks via cosine similarity. This model serves as a simple and interpretable early prototype to facilitate understanding of retrieval behavior without the requirement of any embedding infrastructure support. The retriever returns the top-k chunks used to augment the generation prompt when retrieval is enabled.

\subsection{Prompt Augmentation and Traceability}
The retrieved chunks are inserted into the generation prompt as a separate context block while preserving the user story as the main input. This makes the contextual material supplied to the model available for inspection during validation. The retrieved material provides additional context for generation, but does not by itself establish that the resulting criteria are correct or consistent with the requirements of a particular project.

The retrieval mechanism provides a transparent baseline that allows reviewers to inspect the contextual examples supplied during generation. This visibility supports the examination of retrieval provenance and contextual relevance, but does not establish criterion-level traceability to authoritative project requirements or independently verify the correctness of generated acceptance criteria. Sparse retrieval methods may fail to capture semantically related content expressed using different terminology. Fixed-length token chunks may also separate related constraints across chunk boundaries, making it difficult to retrieve the full context needed for validation. Consequently, the present retrieval configuration should be understood as a prototype baseline rather than a definitive retrieval solution. Future work may explore embedding-based retrieval approaches while preserving the same validation workflow and emphasis on inspectable context.

When RAG is enabled, the interface allows users to inspect the retrieved context during validation. This feature is intended to make grounding visible rather than treating retrieval as a hidden prompt-engineering step. By exposing the contextual material, the system helps users judge whether a generated acceptance criterion is supported by the retrieved contextual examples. If RAG was not enabled for a model, the interface explicitly indicates that no retrieval context was provided.

\subsection{Multi-Model Generation Strategy}
To reduce dependency on a single model and support comparative review, the assistant integrates four models:
\begin{itemize}
    \item \textbf{FLAN-T5-Large }(local inference)
    \item \textbf{Gemini 2.5 Flash }(API)
    \item \textbf{OpenAI gpt-4o-mini} (API)
    \item \textbf{LLaMA-3.3} via Together.ai (API)
\end{itemize}
The tool enforces consistent prompt structure and output formatting across models to support side-by-side comparison.

\section{System Architecture and Implementation}

The prototype described in this paper represents the implementation developed to investigate the proposed validation workflow; its components and configuration may be refined in subsequent empirical studies. The system architecture facilitates the orchestration of multiple LLMs, including locally hosted and API-based models. A common prompt generator can provide consistent directions for various LLMs, and retrieval can be enabled or disabled independently for each selected model. The system allows reviewers to compare outputs from the selected LLMs and request regeneration of candidate criteria as part of the human-in-the-loop validation process.

The retrieval module is model-agnostic and uses lightweight, traceable techniques. Retrieved documents, prompts, and user actions can be examined as part of the validation process. Feedback and final decisions are stored separately from model outputs so that human actions and model-generated artifacts can be analyzed together.

\subsection{Streamlit-Based Web Interface}

\begin{figure}[!h]
    \centering
    \includegraphics[width=\textwidth,height=0.90\textheight,keepaspectratio]{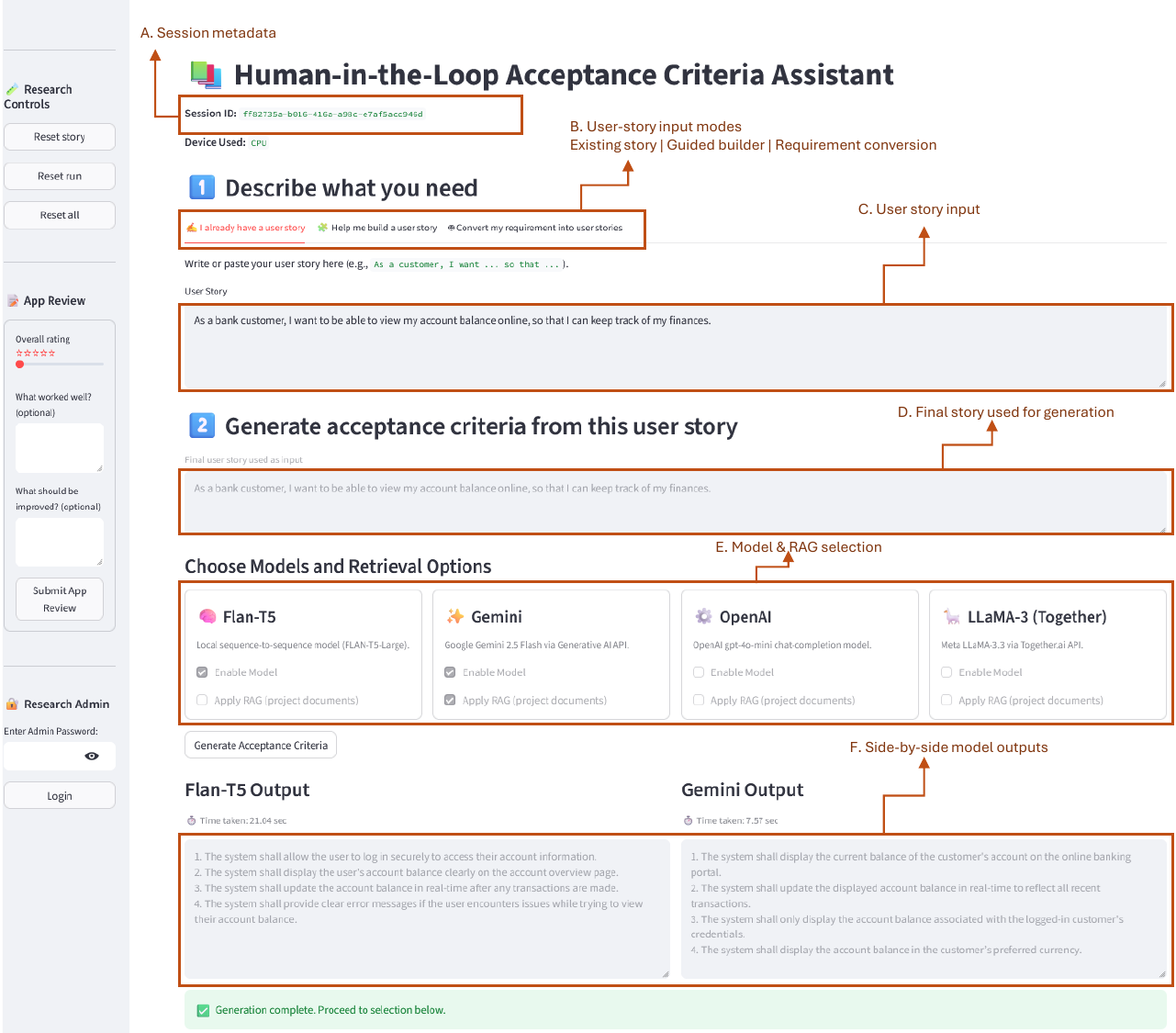}
\caption{Generation and model comparison stage of the Acceptance Criteria Assistant.}
\label{fig:ACAssistantGeneration}
\end{figure}

\begin{figure}[!h]
    \centering
    \includegraphics[width=\textwidth,height=0.90\textheight,keepaspectratio]{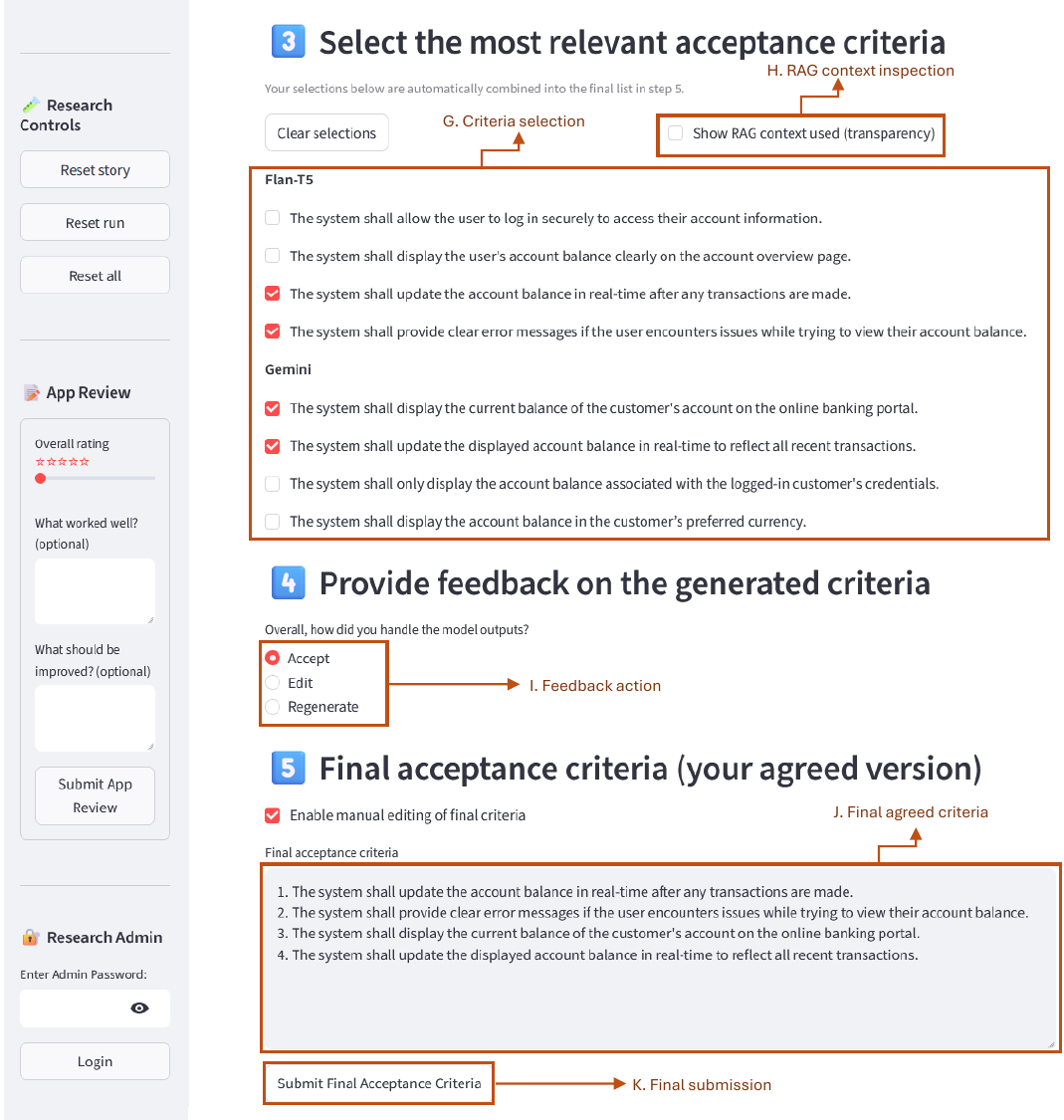}
\caption{Human validation and finalization stage of the Acceptance Criteria Assistant.}
\label{fig:ACAssistantValidation}
\end{figure}

The prototype is implemented as a Streamlit-based web application. Figures~\ref{fig:ACAssistantGeneration} and~\ref{fig:ACAssistantValidation} show the interface as a two-stage workflow. The first stage supports user-story preparation, model and retrieval configuration, and side-by-side generation of acceptance criteria. The second stage supports human validation, including criteria selection, optional inspection of retrieved context, feedback on the generated criteria, and final submission of the agreed version.

The user-story preparation stage supports three input modes. Users may paste an existing user story, construct a user story through a guided form, or provide an informal requirement that the system converts into candidate user stories. This design supports participants with different levels of requirements-engineering expertise while keeping the downstream validation workflow consistent. Before acceptance-criteria generation, the prepared user story is displayed to the user for inspection and, where supported, revision. The user then selects one or more models together with optional retrieval augmentation.

Figure~\ref{fig:ACAssistantGeneration} shows the first stage of the prototype interface. This stage covers user-story preparation, model and retrieval configuration, and side-by-side generation of acceptance criteria. The purpose of this stage is to prepare a clear input artifact and expose the user to alternative model outputs before validation begins.

The annotated elements in Figure~\ref{fig:ACAssistantGeneration} show (A) session metadata, (B) user-story input modes, including existing story, guided builder, and requirement conversion, (C) user story input, (D) the final story used for generation, (E) model and RAG selection, and (F) side-by-side model outputs.

After generation, the workflow moves from model comparison to human validation. The generated acceptance criteria are not treated as final requirements. Instead, they are presented as provisional candidates that the user must inspect, select, edit, or regenerate before submitting the agreed version.

The annotated elements in Figure~\ref{fig:ACAssistantValidation} show (G) criteria selection, (H) optional RAG context inspection, (I) feedback action through accept, edit, or regenerate, (J) reviewer-finalized acceptance criteria, and (K) final submission for storage and later analysis.

Together, Figures~\ref{fig:ACAssistantGeneration} and~\ref{fig:ACAssistantValidation} represent one continuous validation workflow. The user first prepares or selects a user story, configures the models and retrieval settings, and compares generated outputs. The user then validates the generated criteria by selecting relevant items, inspecting retrieved context when available, recording an accept, edit, or regenerate action, and submitting the final agreed criteria. This design makes validation explicit rather than treating AI-generated criteria as automatically accepted.

The example in Figures~\ref{fig:ACAssistantGeneration} and \ref{fig:ACAssistantValidation} presents a banking user story with one primary user goal. It is used to illustrate the generation, comparison, and human validation workflow rather than to define the complexity of requirements supported by the assistant. The prototype is not limited to this example, as it allows users to provide their own user stories or informal requirements and generate candidate acceptance criteria using multiple LLMs. However, the present study does not systematically classify user stories by complexity or evaluate how complexity affects the quality of generated criteria and the effort required for human validation. Future evaluation could examine stories involving multiple conditions, exception scenarios, dependencies, and domain-specific constraints.

The interface is designed to support both comparison and validation. Model outputs are displayed side by side to allow reviewers to compare alternatives rather than relying on a single generated response. Reviewers can then select relevant criteria, edit them if needed, or request regeneration. The workflow records the selected criteria, feedback action, and final submitted result for later analysis.

\subsection{Model Orchestration Layer}

The model orchestration layer builds a consistent prompt for each selected model, attaches retrieved context when enabled, dispatches inference requests, collects responses, and standardizes output formatting for comparison. The study configuration is frozen at the beginning of the generation process so that the selected models, retrieval settings, prompt version, and token settings can be reconstructed during later analysis.

\subsection{Retrieval Module}

The retrieval module includes document storage, a chunking pipeline, TF-IDF vectorization, cosine-similarity ranking, top-k retrieval, and prompt-context construction. The module is separated from the model layer so that future retrieval strategies can be introduced without changing the overall human validation workflow.

\subsection{Logging and Data Capture}
The prototype records generated outputs and instrumented human validation events. In addition to the generated prompts and outputs, the stored data includes the session ID, timestamp data, device information, frozen study configuration, the user story submitted by the participant, models used, retrieval configuration, generated outputs per model, selected criteria, any edited outputs, finalized acceptance criteria, and an event log with information on interaction events. The event log data allows for tracking of key validation activities for each run. It records configuration locking, generation initiation and completion, selection updates, changes to edited outputs, and final submission events. The final submission is saved to a private research repository for later analysis.

The logging design distinguishes intermediate reviewer interactions from the final submitted artifact. Selection changes and text revisions describe how the candidate criteria are modified during review, whereas final submission identifies the reviewer-finalized result. This distinction allows subsequent analysis to examine the validation process separately from its final outcome. The system also includes an optional app review form that collects a rating and open-ended feedback about what worked well and what should be improved. These app-level reviews are not acceptance-criteria validation data, but they provide useful usability feedback for improving the research prototype.

Before using the assistant, participants completed a research participation form and provided affirmative consent for research data collection. Name, email address, and affiliation were optional, while self-reported IT and requirements-engineering experience information was required.

Participation records were stored separately from the validation logs, although the records remained linkable through session identifiers. The collected research data were stored in a private, access-controlled repository. Identifying information was excluded from the aggregate reporting of interaction patterns.

This logging allows for future research into various aspects of retrieval-based validation, including acceptance rates, edit behavior, time needed for final decision, level of agreement between different models, and the relation of participant background to validation actions.

\section{Human-in-the-Loop Validation Workflow}
The tool operationalizes HITL validation as a structured decision process:
\begin{enumerate}
    \item \textbf{Prepare:} enter an existing user story, create a user story using a guided form, or convert an unstructured requirement into user stories.
    \item \textbf{Generate:} produce acceptance criteria from one or more selected models, with or without retrieval support for each model.
    \item \textbf{Compare:} show outputs side-by-side to support cross-checking.
    \item \textbf{Inspect:} display retrieved context when RAG was enabled, allowing the user to examine the evidence provided to the model.
    \item \textbf{Decide:} the user can select one of three interface actions:
\begin{itemize}
    \item \textbf{Accept:} retain the current selected acceptance criteria without an additional manual text edit;
    \item \textbf{Edit:} manually revise the criteria before finalization;
    \item \textbf{Regenerate:} request new candidate criteria for further review.
\end{itemize}
    \item \textbf{Finalize:} produce and store the agreed acceptance criteria.
    \item \textbf{Log:} record the interaction trace for analysis.
\end{enumerate}

The workflow records the selected criteria, the feedback state, and the final submission. These records make the reviewer interventions visible for later analysis. A final Accept state does not necessarily mean that all generated criteria were retained, since the reviewer may have selected only a subset before submission. In this paper, ratification refers to the recorded final submission of the reviewer-selected result. It makes the finalization decision observable, but it does not independently establish the correctness of the criteria or agreement among all relevant stakeholders.

The current prototype presents criteria from different model panels separately to preserve their source provenance and support comparison. However, different models may generate criteria that describe similar expected behavior, increasing the amount of text reviewers need to inspect. For example, Figure~\ref{fig:ACAssistantValidation} shows similar criteria concerning account-balance updates after transactions selected from two model panels. The example in Figure~\ref{fig:ACAssistantValidation} illustrates the current selection and finalization workflow rather than an independently assessed set of high-quality acceptance criteria. The retention of similar criteria in the reviewer-finalized result also demonstrates that final submission does not necessarily eliminate redundancy or ambiguity. These quality aspects require separate assessment and motivate further investigation of human validation behavior.

A possible extension would group potentially similar criteria across model panels while preserving the original wording and source-panel information for each variant. The interface could display a representative criterion for each group while allowing reviewers to inspect and select individual alternatives. Such grouping should not automatically merge or discard criteria, because small differences in conditions, constraints, or expected behavior may be important. Reviewers would remain responsible for selecting and revising the final criteria. Whether grouping reduces review effort without hiding meaningful differences would require empirical evaluation.

\section{Preliminary Observations and Planned Evaluation}

The prototype demonstrates that the proposed validation workflow can be implemented as a functioning interactive system and that it can systematically capture the information needed for later empirical analysis. The system supports user-story preparation, optional retrieval grounding, multi-model generation, human selection and editing of acceptance criteria, explicit feedback actions, and final submission of the agreed criteria. In this sense, the contribution of the current work is not a completed evaluation of model quality, but an instrumented platform for studying how validation of AI-generated requirements artifacts takes place.

The implementation illustrates several design characteristics of the validation workflow. The three input modes support different ways of preparing a user story. Side-by-side model outputs allow reviewers to compare alternative criteria before finalization. Retrieved context can be inspected when enabled, making the information supplied during generation visible to the reviewer. Criterion-level selection, editing, feedback states, and final submission provide observable interaction records for later analysis.

Further empirical evaluation can examine four main dimensions of the validation workflow. The first is intervention frequency, namely how often users accept, edit, or regenerate outputs under different model and retrieval settings. The second is the nature of human edits, including removal of unsupported criteria, addition of missing constraints, clarification of ambiguous language, and reformulation into more testable criteria. The third is validation effort, including the number of selected criteria, manual edits, and time or interaction steps required to reach a final agreed version. The fourth is model disagreement, especially whether disagreement between models predicts higher rates of human revision.

Therefore, we do not claim that the prototype already improves correctness or eliminates hallucination. Instead, it provides the workflow and data collection infrastructure needed to investigate those questions systematically in future studies.

\section{Discussion and Implications for Verification}

The proposed approach reflects a broad interpretation of verification and validation in requirements engineering. Acceptance criteria are not executable code, but they strongly influence downstream development and testing because they define expected system behavior. When such criteria are generated by AI, their fluency does not guarantee correctness, completeness, or alignment with stakeholder intent. Validation must therefore involve contextual grounding, human review, and an explicit decision about whether the generated text is acceptable.

Retrieval contributes to the proposed workflow by supplying additional contextual material during generation and making that material available for human inspection. Reviewers can examine the retrieved context when assessing generated criteria. However, retrieved examples should not be treated as authoritative evidence of project-specific correctness. The current implementation therefore supports inspection of the generation context rather than automatic verification of the generated acceptance criteria.

The human-in-the-loop workflow provides the second component of the validation process. Criteria generated by the models are treated only as candidate criteria that need to be selected, edited, regenerated, or ratified by a user before being finalized. This ratification stage is central to the system, because validation is not hidden as an informal correction activity but is represented as an observable decision process.

The multi-model workflow also contributes to validation. Different models may emphasize different aspects of the same user story, and their outputs may reveal ambiguity, lack of context, or possible over-specification. Thus, differences between model-generated criteria are not only a nuisance but also a useful indicator that closer human inspection may be needed.

Overall, the proposed system shows that trustworthy AI support in requirements engineering does not depend only on generating better criteria through more capable models. It also requires interfaces and logging mechanisms that make generated artifacts inspectable, human decisions explicit, and validation behavior analyzable. Instead of verifying generated artifacts through mathematical proofs or automated testing, the proposed method supports validation through retrieval grounding and human assessment. This is appropriate because requirements engineering also relies on assessment, interpretation, and stakeholder validation \cite{nuseibeh2000requirements}.

\section{Conclusion and Future Work}
\label{sec:conclusions}

This paper presented a validation-oriented method for AI-generated acceptance criteria in requirements engineering. The main idea of this work is not only to generate acceptance criteria from user stories, but to support their validation before they are accepted as final requirements artifacts. For this purpose, the proposed approach combines retrieval-based grounding, multi-model generation, and human-in-the-loop validation.

The prototype shows how this idea can be implemented in practice. The system allows users to prepare or enter a user story, select one or more models, optionally enable retrieval augmentation, compare generated acceptance criteria side by side, inspect retrieved context, and then accept, edit, or regenerate the generated criteria. The reviewer-finalized acceptance criteria are stored together with the interaction data, which makes it possible to analyze how humans interact with AI-generated requirements artifacts.

The contribution is a working, instrumented validation workflow rather than an empirical evaluation of model accuracy. Because fluent AI-generated criteria may still be unsupported, incomplete, or hallucinated, the approach makes retrieved context inspectable and requires human review and explicit finalization.


Future research will focus on the empirical analysis of the collected data. In particular, we plan to examine how retrieval configurations, model choice, and user-story complexity influence human validation decisions. We also plan to analyze the types of edits made by users, the frequency of accept, edit, and regenerate actions, and whether retrieval augmentation affects the occurrence of unsupported acceptance criteria, using independent assessments of generated requirements quality. Another direction is to investigate provenance-preserving grouping of similar criteria generated by different models. Future studies should examine whether such grouping reduces review effort while retaining important differences between candidate criteria.

A further direction is to extend the prototype from validating generated acceptance criteria to identifying requirements gaps across related software artifacts. In addition to checking whether generated criteria are grounded and acceptable, future versions could support gap-oriented analysis, including missing requirements, hidden assumptions, inconsistencies, decision voids, traceability gaps, and stakeholder silence gaps. This would broaden the role of the system from acceptance-criteria validation toward a more general form of AI-assisted requirements quality analysis.

\subsubsection{\ackname} This work has been supported by FAST, the Finnish Software Engineering Doctoral Research Network, funded by the Ministry of Education and Culture, Finland. Support from Business Finland is also gratefully acknowledged (Projects ANSE (1822/31/2025) and GENIUS (2545/31/2024)).



\bibliographystyle{splncs04}
\bibliography{References}
\end{document}